\documentclass[]{aa} 
\usepackage{txfonts,epsfig,graphicx,natbib,url,twoopt,amsmath,algorithm2e}
\usepackage{color}

\DeclareMathOperator*{\argmin}{arg\,min}

\usepackage{natbib,twoopt}
\bibpunct{(}{)}{;}{a}{}{,}             
\makeatletter
  \newcommandtwoopt{\citeads}[3][][]{\href{http://adsabs.harvard.edu/abs/#3}%
    {\def\hyper@linkstart##1##2{}%
     \let\hyper@linkend\@empty\citealp[#1][#2]{#3}}}
  \newcommandtwoopt{\citepads}[3][][]{\href{http://adsabs.harvard.edu/abs/#3}%
    {\def\hyper@linkstart##1##2{}%
     \let\hyper@linkend\@empty\citep[#1][#2]{#3}}}
  \newcommandtwoopt{\citetads}[3][][]{\href{http://adsabs.harvard.edu/abs/#3}%
    {\def\hyper@linkstart##1##2{}%
     \let\hyper@linkend\@empty\citet[#1][#2]{#3}}}
  \newcommandtwoopt{\citeyearads}[3][][]%
    {\href{http://adsabs.harvard.edu/abs/#3}
    {\def\hyper@linkstart##1##2{}%
     \let\hyper@linkend\@empty\citeyear[#1][#2]{#3}}}
\makeatother

\begin{document}

\title{Sparse inversion of Stokes profiles}
\subtitle{I. Two-dimensional Milne-Eddington inversions}

\author{A. Asensio Ramos\inst{1,2} \& J. de la Cruz Rodr\'{\i}guez\inst{3}}

\institute{
 Instituto de Astrof\'{\i}sica de Canarias, 38205, La Laguna, Tenerife, Spain; \email{aasensio@iac.es}
\and
Departamento de Astrof\'{\i}sica, Universidad de La Laguna, E-38205 La Laguna, Tenerife, Spain
\and
Institute for Solar Physics, Dept. of Astronomy, Stockholm University, Albanova University Center, 10691 Stockholm, Sweden
\email{jaime@astro.su.se}
}
             
  \date{Received ---; accepted ---} 

  \abstract
  {Inversion codes are numerical tools used for the inference of physical properties from the observations. Despite their success, the quality of 
  current spectropolarimetric observations and those expected in the near future presents a challenge to current inversion codes.}
  {The pixel-by-pixel strategy of inverting spectropolarimetric data that we currently utilize needs to be surpassed and improved.
  The inverted physical parameters have to take into account the spatial correlation that is present in the data
  and that contains valuable physical information.}
  {We utilize the concept of sparsity or compressibility to develop an new generation of inversion codes for the 
  Stokes parameters. The inversion code uses numerical optimization techniques based on the idea of proximal algorithms to
  impose sparsity. In so doing, we allow for the first time to exploit the presence of spatial correlation 
  on the maps of physical parameters. Sparsity also regularizes the solution by reducing the number of unknowns.}
  {We compare the results of the new inversion code with pixel-by-pixel inversions, demonstrating the increase in
  robustness of the solution. We also show how the method can easily compensate for the effect of the telescope 
  point spread function, producing solutions with an enhanced contrast.}
  {}

   \authorrunning{Asensio Ramos \& de la Cruz Rodr\'{\i}guez}
   \titlerunning{Sparse inversion of Stokes profiles}
   \maketitle
%

\section{Introduction}
An inversion code is a computer program based on algorithms that allows the user to 
extract information about the parameters of a physical
system from the interpretation of observables\footnote{The term \emph{inversion} comes from
the fact that, given a forward model defined by an operator $\mathbf{O}$, one needs to apply 
the (nonlinear) inverse of the operator, $\mathbf{O}^{-1}$, to the observations to recover the model parameters.
From a strict point of view, it would be better to consider these algorithms as \emph{inference}
codes, although we stick with the classical term inversion to avoid confusion.}.
In solar physics, nonlinear inversion 
codes are routinely applied after their development in the '70s to get the thermodynamical and 
magnetic properties of different regions of the solar atmosphere from the 
interpretation of the Stokes parameters \citep{harvey72,auer_heasly_house77,skumanich_lites87}.

One of the fundamental problems in general, and specifically for spectropolarimetric inversions,
is that the relation between the observables and the physical parameters of interest is
very convoluted. Despite the strong nonlinearity and nonlocality inherent to the
radiative transfer, it has been possible to develop some simple diagnostics during the past years:
the line ratio technique \citep{stenflo73,stenflo10,stenflo11}, the center-of-gravity method 
\citep{semel70,rees_semel79} and calibration curves between spectral
line summaries and certain magnetic field properties \citep[e.g.,][for recent applications]{lites08,imax11}.

The appearance of the first powerful computers allowed researchers to apply optimization
techniques for nonlinear functions and use more elaborate models. A $\chi^2$ merit 
function\footnote{That is a direct consequence of the assumption that the observations
are corrupted with additive Gaussian noise.} is minimized with respect to the physical
parameters defining the specific model. The first efforts (with the application of 
the computers available at that time) made use of the Milne-Eddington (ME) approximation 
to analytically solve the radiative transfer equation \citep[e.g.,][]{harvey72,auer_heasly_house77,landi_landolfi04}. Although the
simplifying assumptions that one needs to use when using the ME approximation may not be fully fulfilled in real
solar plasmas, it is still one of the most widely used models, in part because of its
simplicity. This simplicity leads to very fast inversion codes that can be applied to the 
large number of observations that we currently obtain. State-of-the-art inversion codes such as VFISV \citep{borrero07,borrero_vfisv10},
used for inferring magnetic field vectors from the Helioseismic and Magnetic Imager (HMI; onboard
the Solar Dynamics Observatory) data, MILOS \citep{orozco_hinode07} and
MERLIN \citep{skumanich_lites87,lites07}, currently applied to data from the Hinode spacecraft, or
the codes based on look-up tables and the principal component analysis (PCA) decomposition \citep{rees_PCA00,socas_arturo_lites01} used
for the inversion of THEMIS (T\'elescope h\'eliographique pour l'\'etude du magn\'etisme et des instabilit\'es solaires) data, are based on the ME approximation.

The availability of more powerful computers in recent years allowed us to use more
complex and more realistic models. One of the essential ingredients of this 
revolution was the application of the idea of response functions \citep{landi_response77} 
to the inversion of Stokes profiles with non-trivial
depth stratifications of the physical quantities. The first representative
of this family of codes was SIR \citep[Stokes Inversion based on Response functions;][]{sir92}.
Such evolution occured naturally at that time because observations were showing strong asymmetries in the
Stokes profiles in magnetized regions. The explanation of such asymmetries requires the presence of 
gradients along the line-of-sight (LOS) of the physical properties \citep{solanki88,grossman_doerth88,solanki93,sigwarth99,lopezariste02,khomenko03,marian08,viticchie_2_11}.
Another representative of these inversions code is SPINOR \citep{frutiger00}.
Based on the same strategy, \cite{socas_trujillo_ruiz00} developed the NICOLE code \citep{socas_nicole14}, capable of dealing
with lines in NLTE (non-local thermodynamical equilibrium). This model
has been mainly applied for the inversion of Ca \textsc{ii} infrared triplet
lines, which are formed under strong NLTE conditions \citep[e.g.,][]{socas_trujillo_ruiz00,jaime12}.

Another step forward in the field of inversion codes was carried out by \cite{asensio_martinez_rubino07} and \cite{asensio_hinode09},
who introduced Bayesian inference for spectropolarimetric observations. This allows the user
to obtain posterior probability distributions for any model parameter and their correlation
with the remaining ones. This probabilistic inference is extremely powerful but requires a
huge effort in terms of computational power.


Arguably, the latest step in the field of inversion codes has
been to compensate the effect of the telescope point spread function
(PSF)\footnote{The telescope PSF smears the intensity of point
sources spatially, mixing information from neighbouring features.} 
that is present in the observed data. \cite{vannoort12}
implemented a forward approach, where synthetic spectra from the
inversion are convolved spatially with the telescope PSF before comparison
with the observations. \cite{ruizcobo_asensioramos13} used a regularized
deconvolution to compensate the observations before the inversions
were performed. 

The former solves the full problem in which the spatial convolution and the
inversion are coupled in the same operator $\mathbf{O}$, resulting in a very large-scale classical 
inversion code. The use of a Levenberg-Marquardt (LM) scheme \citep{levenberg44,marquardt63}
with a compact PSF leads to a sparse Hessian matrix that largely
facilitates the inversion by spatially coupling the convergence of adjacent pixels. On the contrary, the latter
decouples the inversion. First the observations are deconvolved by applying a PCA 
regularization in the spectral domain and the application of a standard Richardson-Lucy \citep{richardson72,lucy74}
maximum likelihood deconvolution for each eigenimage. Then, the resulting denoised and deconvolved data are inverted using
any classical inversion code.  Both approaches introduce different regularizations in the inversion problem and can lead to
potentially different results. 

Also, in the former approach there is no need to consider a low-pass filtering or a
regularization of the deconvolution, but the author reports cases of oscillatory patterns in
the inferred model that need to be damped during the inversion. In
practice, the implementation of the second approach is easier, cleaner and
less prone to the aforementioned oscillatory artifacts, but the PCA
regularization can filter out features with low statistical weight in
the dataset. These methods rely upon accurate knowledge of the PSF used in the
inversion, which may not be always available or applicable.

We note that \cite{2013A&A...553A..63S} also combined data inversions with
a regularized deconvolution to compensate
ground-based observations for straylight, believed to originate from
small-scale aberrations \citep{2012A&A...537A..80L}. 

To take a new step forward in the field of inversion codes, in this
paper we propose a new technique to spatially couple the
parameters of the model, inspired on the idea of \emph{sparsity or
compressibility} of data. This method allows to use a PSF in the
same way as \cite{vannoort12} but it is not strictly needed for the
spatial coupling.

\section{Sparsity regularization}
\label{sec:sparsityMethod}
With the use of fast slit spectropolarimeters and also two-dimensional 
filterpolarimeters, we routinely have 2D maps of regions in the solar 
atmosphere with the four Stokes parameters observed at several points along one or 
several spectral lines. This rate of new high-quality 2D observations will
increase in the future with the advent of bi-dimensional spectropolarimeters based on
image slicers or optical fibers. The interpretation of these observations have been done in the past by assuming that 
all pixels are completely unrelated and applying the inversion codes in a 
pixel-by-pixel basis. After this pixel-by-pixel inversion, the spatial
smoothness of the derived quantities is taken as an 
indication of the success of the inversions. Salt-and-pepper noise 
present in the inverted maps of physical parameters is an indication
of problems: either the required information cannot be extracted
from the Stokes profiles, or the inversions failed to converge to
a good solution.

It has become more and more evident that two-dimensional observations
and the ensuing inversions are needed to fully understand the physical
processes in the solar atmosphere. This is the case even when one assumes local thermodynamical 
equilibrium (LTE), which relates radiation to the local
properties of the plasma. The first representatives of such
approach are the already described codes of \cite{vannoort12} and \cite{ruizcobo_asensioramos13}, where the
PSF of the telescope couples the observed Stokes profiles of nearby pixels.

\subsection{The general idea}
Here we develop the general idea of regularized inversion codes, with the
regularization being based on the idea of \emph{sparsity}. 
Sparsity or compressibility idealizes the concept that the data can
be projected to a parameter space where a \emph{reduced} set of variables
can be used to fully describe that dataset.
The motivation for this regularization resides on a very simple observation. When one
saves a continuum image as a raw file (for instance, a standard 512$\times$512 pixel image), 
the size of the file is roughly 1 MB (using 4 bytes per pixel). The same image compressed
using a lossless file format reduces the size by a factor 3-4, while a lossy format can go
further and increase the compression ratio to a factor $\sim 10$. The compression
is possible, fundamentally, due to the existence of spatial correlation (properties like smoothness,
structures in the image, \ldots) on the image. If appropriately exploited, 
it is possible to predict the value of a certain pixel from the values of other pixels thus making it unnecessary 
to store the value of all pixels.


Let us assume that we have observed the Stokes parameters on a 2D grid of $N_\mathrm{pix}=N_x \times N_y$
pixels for a set of $N_\lambda$ wavelength points that sample one or several spectral lines. 
We propose a model atmosphere for every pixel to explain the observations, each model atmosphere
being defined with a set of $N_\mathrm{par}$ parameters, so that each pixel $k$ is described with the
vector of parameters $\mathbf{\hat{p}}_k=(p_{1k}, p_{2k}, \ldots, p_{N_\mathrm{par}k})$.
We can build the vector of parameters $\mathbf{p}$ that encode all the model parameters for
all the pixels in the field-of-view, such that $\mathbf{\hat{p}}=(\mathbf{\hat{p}}_1,\mathbf{\hat{p}}_2,\ldots,\mathbf{\hat{p}}_{N_\mathrm{pix}})$.
It is possible to reorder this vector so that it is built by stacking the value of each physical parameter for all the pixels, so that
$\mathbf{p}=\{\mathbf{p}_1,\mathbf{p}_2,\ldots\,\mathbf{p}_{N_\mathrm{par}}\}$, where $\mathbf{p}_i$ encodes the value of a certain parameter (i.e., magnetic field inclination, Doppler
velocity, etc.) as a 2D map. Note that the vectors $\mathbf{p}$ and $\mathbf{\hat{p}}$ are equivalent except for the ordering.
The aim of any inversion code is, then, to infer the full vector $\mathbf{p}$ from the 
observed Stokes profiles. As explained in the introduction, classical inversion codes deal with this
problem by inverting the observed Stokes profiles pixel by pixel. 

In our approach, instead of working directly on the real space of parameters, we use a
linear transformation so that the $i$-th parameter is given by:
\begin{equation} 
\mathbf{q}_i = \mathbf{W}[\mathbf{p}_i].
\label{eq:linearTransformation}
\end{equation}
In the previous equation, $\mathbf{W}$ is a linear operator that transforms from the real space to 
the transformed space. Among many potential transformations, the most interesting in our context are the Fourier, 
wavelet or discrete cosine transforms. It is clear that by working on the transformed space and using an inversion
code to infer the transformed parameters we have gained nothing. However, here we can impose a
fundamental key ingredient: \emph{the assumption that the transformed image 
is sparse}\footnote{It is more precise to say that the physical parameters are \emph{compressible} in the 
transformed domain, which means that the majority of coefficients in the transformed space are small.} \emph{in the transformed 
domain}. In other words, if the appropriate transformation $\mathbf{W}$ is used, 
many elements of the transformed image (that we term modes in the following for simplicity) are zero or very close to zero in absolute value. 
Sparsity, that is usually fulfilled in nature, has been shown to be an extremely powerful assumption. Techniques like
compressed sensing \citep{candes06,donoho06} or exact matrix completion from partial measurements \citep{candes_recht08}
strongly rely on the sparsity assumption.

\subsection{Testing for sparsity in physical parameters}
For the inversion that we describe in the following to work properly, it is crucial to test for the
sparsity of the maps of model parameters. To this end, we do ME pixel-by-pixel inversions of two maps
observed with the Solar Optical Telescope/Spectro-Polarimeter SOT/SP \citep{lites_hinode01} aboard 
\emph{Hinode} \citep{kosugi_hinode07}. We choose to only invert the Fe \textsc{i} line at 6302.5 \AA.
The first map is a quiet Sun map extracted from the observations analyzed 
by \cite{lites08}, which were obtained at disk center on 2007 March 10. The second map is a sunspot 
of the active region NOAA 10953, that was mapped at an average heliocentric angle of 
$\theta=12.8^\circ$ on 2007 April 30. 
This inversion is carried out with a classical inversion code
using the LM algorithm and no regularization whatsoever, but the
inversion is repeated 10 times for each pixel using different initial
values of the model parameters (which are randomized). The ME model depends on a set of
assumptions: i) the atmosphere is plane-parallel, semi-infinite, and in local thermodynamic equilibrium, ii) the line opacity, 
Doppler width, and all line properties are constant with depth on the atmosphere, iii) the magnetic field vector is constant with 
depth in the atmosphere, and iv) the Planck function is a linear function of the continuum optical depth $\tau_c$, so that 
$B_T=B_0+B_1 \tau_c$, where $\mathrm{d}\tau_c=-\kappa_c \mathrm{d}s$, where $\kappa_c$ is the continuum opacity. The ME
model parameters of our code are: Doppler width of the line in wavelength units ($\Delta \lambda$), macroscopic 
bulk velocity ($v_\mathrm{mac}$), the slope $B_1$ and intercept $B_0$ of the source function, the ratio between the line 
and continuum absorption coefficients ($\eta$), the line damping parameter ($a$) and the magnetic field vector parameterized 
by its modulus, inclination and azimuth with respect to a given reference direction ($B$, $\theta_B$ and $\phi_B$, respectively).

\begin{figure*}
\centering
\includegraphics[width=1.0\textwidth, trim = 0.6cm 0 0.5cm 0, clip]{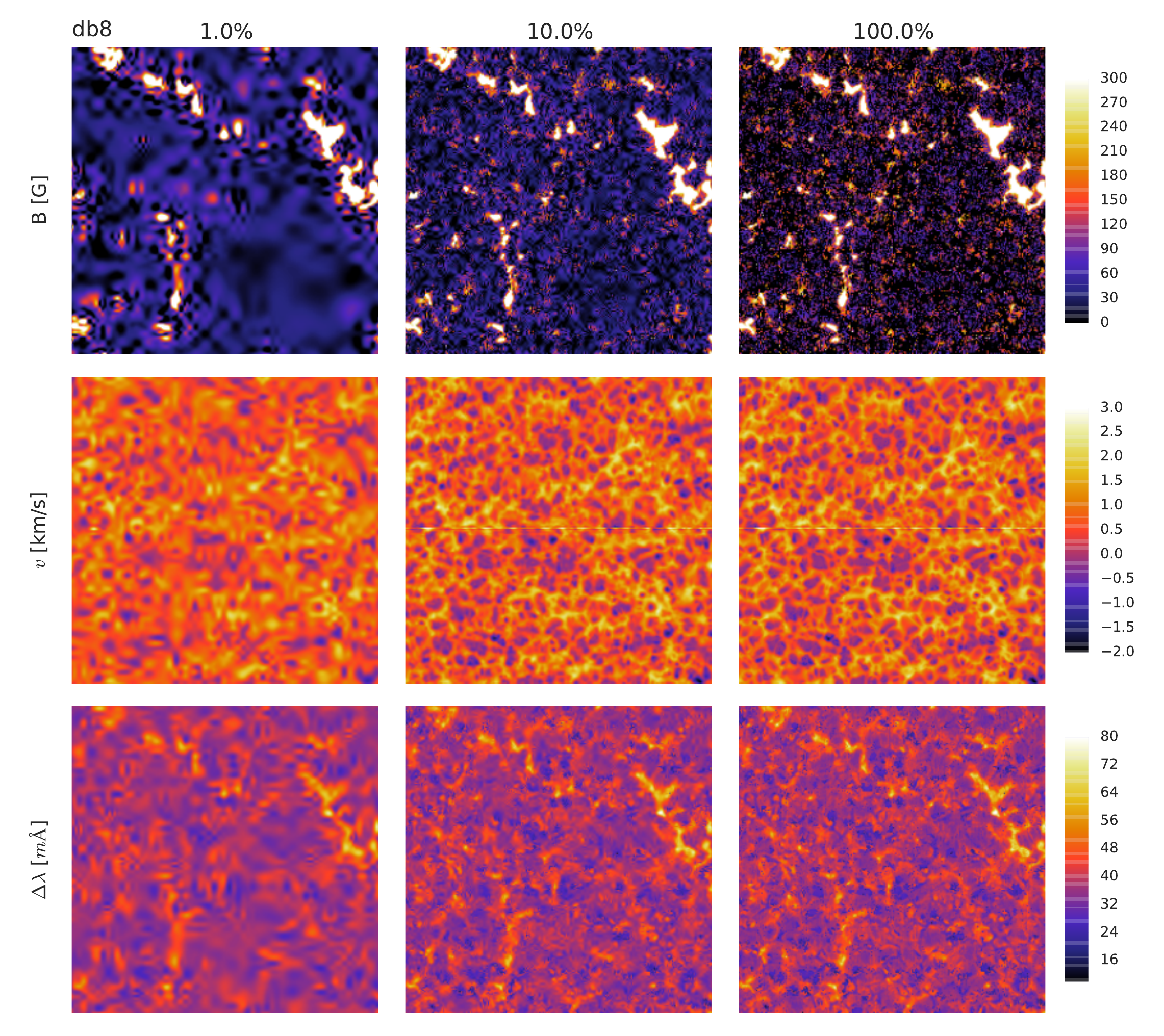}
\caption{Test for the compressibility of the parameters inferred in a quiet Sun map. Each column indicates the number of non-zero coefficients of the original map
that are retained.}
\label{fig:testSparsityQuiet}
\end{figure*}

\begin{figure*}
\centering
\includegraphics[width=1.0\textwidth, trim = 0.6cm 0 0.5cm 0, clip]{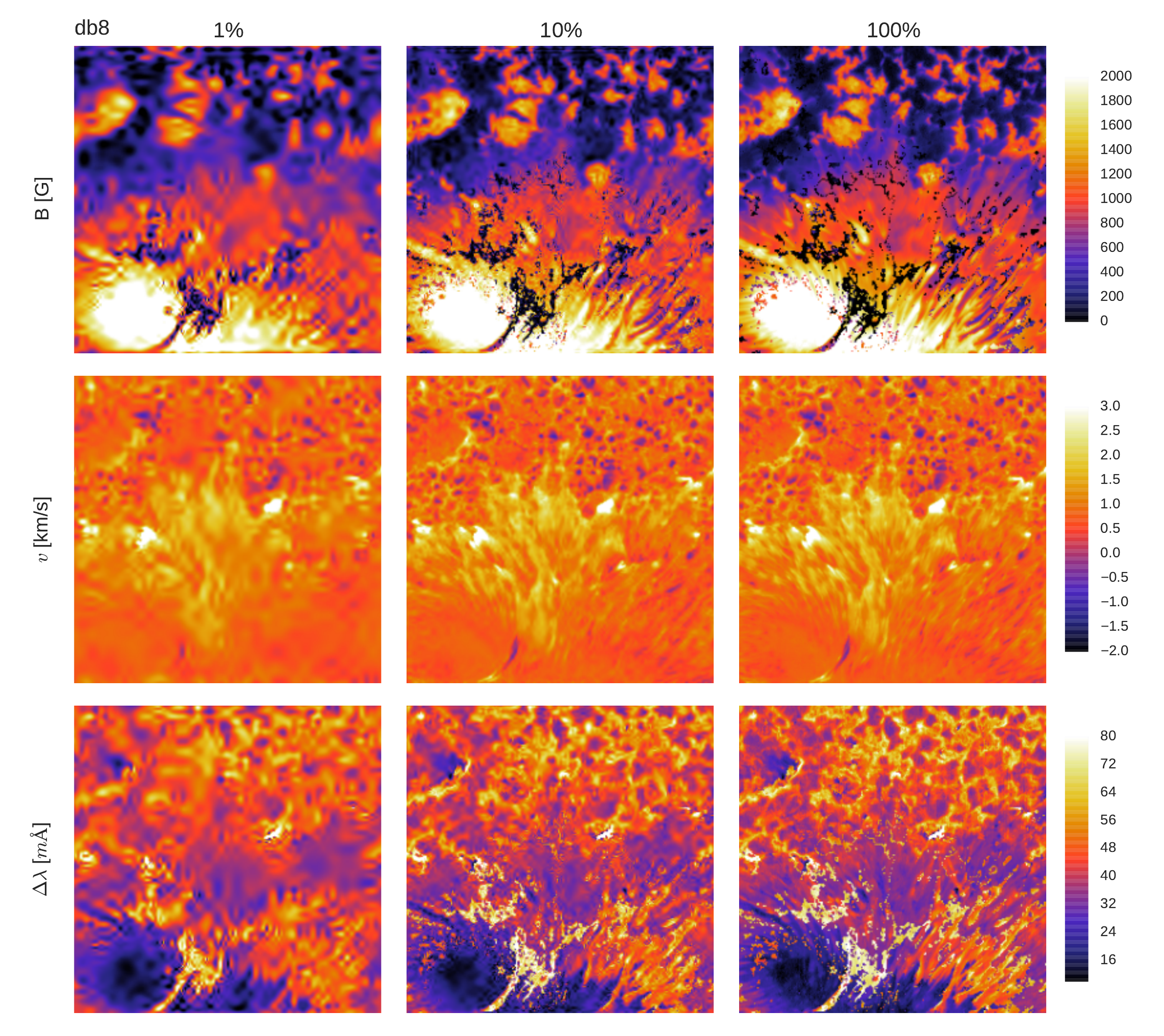}
\caption{Same as Fig.~\ref{fig:testSparsityQuiet} but for the sunspot data. We note that these images correspond to the solution of the pixel-by-pixel inversion and contain some artifacts.}
\label{fig:testSparsitySpot}
\end{figure*}

\begin{figure*}[!ht]
\centering
\includegraphics[width=1.0\textwidth, trim=0.0cm 0 0.0cm 0, clip]{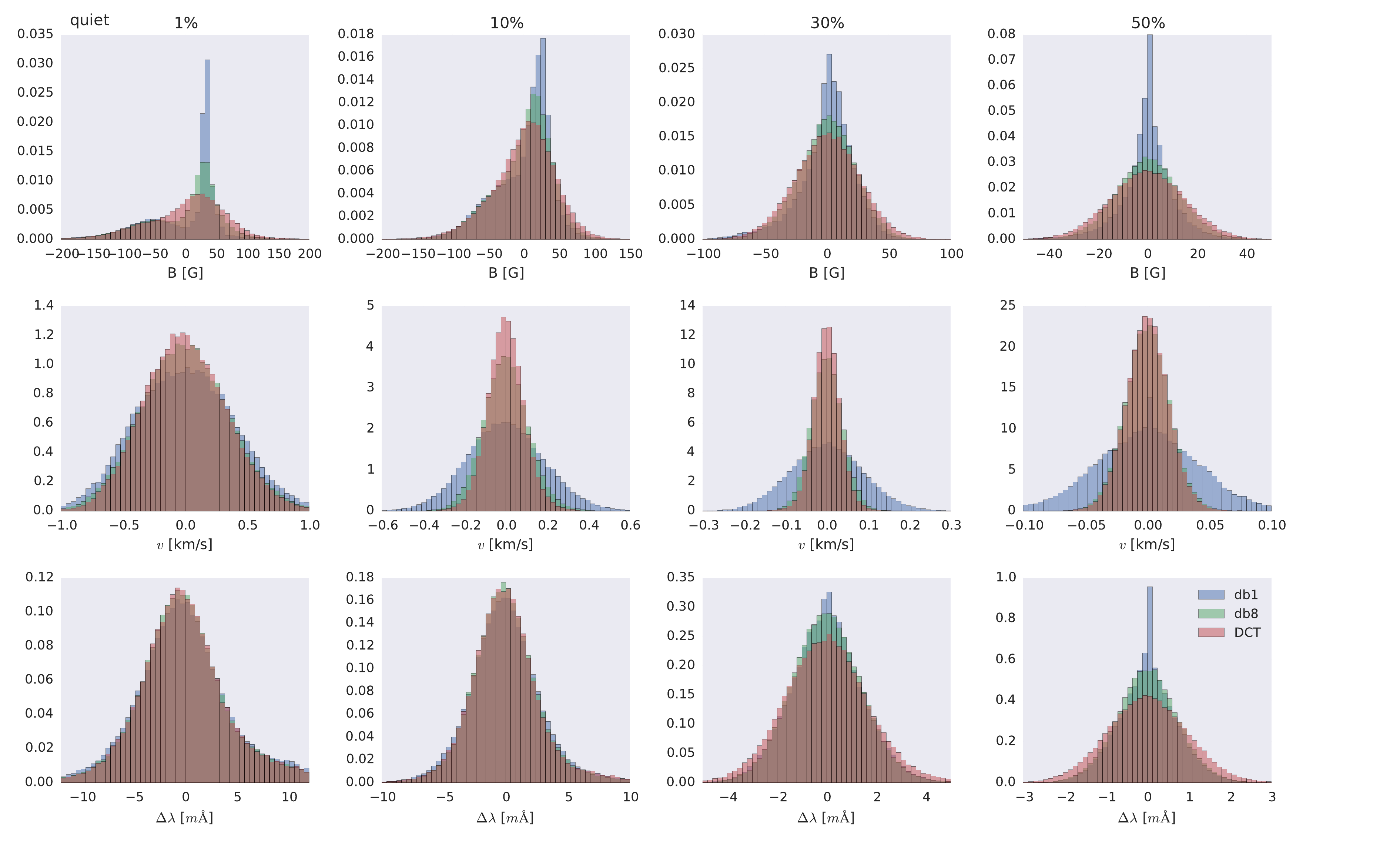}
\includegraphics[width=1.0\textwidth, trim=0.0cm 0 0.0cm 0, clip]{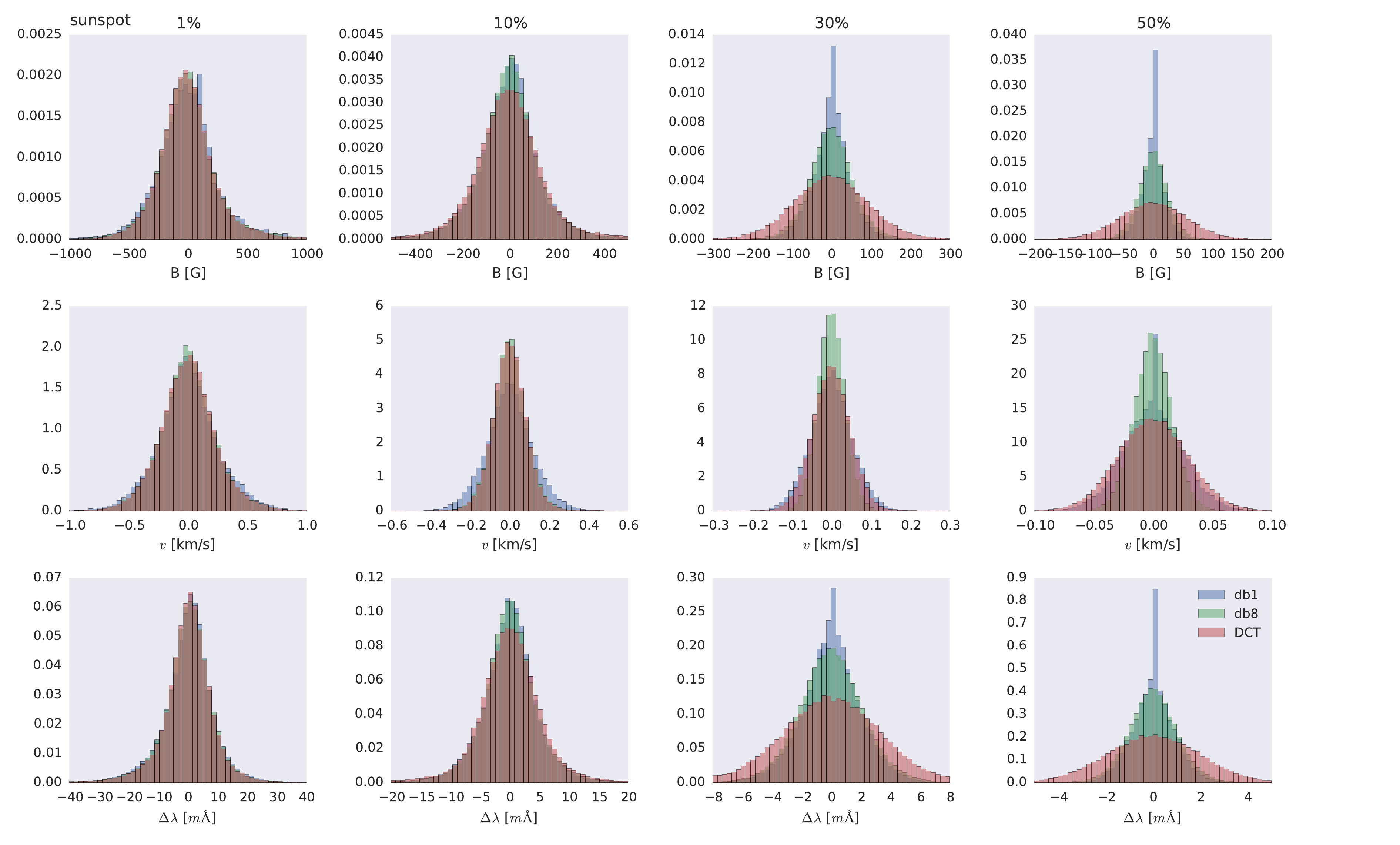}
\caption{Histograms of the difference between the original parameters and the compressed ones at different levels (indicated in each column).
The upper panels show the results for the quiet Sun map, while the lower panels refer to the sunspot data. Each histogram refers to a different
linear transformation.}
\label{fig:testSparsityDiff}
\end{figure*}

Figures \ref{fig:testSparsityQuiet} and \ref{fig:testSparsitySpot} display the effect of compressing some parameters in the quiet Sun and the sunspot maps. Both maps contain 
256$\times$256 pixels and the selected parameters are the magnetic field strength, Doppler velocity and 
width of the line. Each panel shows the reconstruction of the original image (the rightmost image in each panel) when the image is transformed using a 
linear transform, thresholded leaving
only a certain percentage of the largest coefficients and transformed back. The percentage of non-zero coefficients is shown as a label in each
column. In this example, we use the Daubechies-8 (db8) orthogonal wavelet \citep{ripples01} as a sparsity enhancing transform, which is appropriate for smooth maps, although
the results are similar if other transforms are used.
The figure demonstrates that one gets a very well reproduction of the original maps when the number of non-zero
coefficients is slightly larger than 15\%. This compression can be obtained because the linear transformations capture part of the spatial correlation in
the image. As a consequence, the value of $\sim$85\% of the pixels can be predicted from the value of just 15\% of the image plus the presence
of spatial correlation. 
To help in this comparison, Fig. \ref{fig:testSparsityDiff} displays the histogram of the difference
between the original magnetic field strength, Doppler velocity and width of the line and the compressed one for different levels of
compression. In this case, apart from the Daubechies-8 orthogonal wavelet transform, we also introduce the Haar (db1) orthonormal wavelet \citep[equivalent to the Daubechies-1 wavelet;][]{ripples01}, that
is a discontinuous wavelet appropriate for representing efficiently maps with spatial discontinuities and the discrete cosine transform (DCT), appropriate
for periodic signals.
At the light of these results, keeping 10-15\% of the coefficients seems to reproduce very well the
original maps of parameters. When the fraction of remaining coefficients is made too small, more artifacts appear and the difference between the original
maps and the compressed ones are too large.



One of the consequences of keeping only a certain percentage of the coefficients in the transformed domain is that
we loose some information (lossy compression). It is important that the user of the inversion code described in this paper selects
appropriate threshold limits so that no important information is lost in the final maps. Although we recognize the presence of these 
additional parameters to tweak, standard inversion codes are not easy to use and are also full of parameters to tweak. At the end,
all these parameters can be tuned during an initial trial-and-error phase and fixed afterwards.

\subsection{Advantages and disadvantages}
The inversion code that we describe in the following (together with the numerical techniques
needed to enforce sparsity) infers the physical parameters in the 
transformed space. At the end, the physical parameter of interest is obtained by just a final application of the inverse transform.

Working on the transformed domain and applying a sparsity regularization leads to the following obvious advantages:
\begin{itemize}
\item Imposing that the solution to the inversion has to be sparse introduces a strong regularization because 
the number of unknowns is heavily reduced. From the potential $N_\mathrm{par} N_x N_y$ to only
$s N_\mathrm{par} N_x N_y$, with $s \ll 1$.
We have found that typical values of the sparsity $s$ will be around 20-30\%.
Given that a large fraction of the coefficients can
be set to zero without degrading the solution, the number of free
variables that we are currently using in pixel-by-pixel inversions is
highly overestimated. An overly large number of degrees of freedom can produce
nonphysical fluctuations of the physical parameters\footnote{We
note that the response functions must be computed for every
parameter, despite the assumption of sparsity, to assess which
coefficients are negligible.}.
\item The inverse linear transformation gives the value of a physical parameters at a specific
observed pixel as a linear combination of all modes in the map. This global character induces changes in all
the pixels of the original space simultaneously when perturbing the value of a single mode. Reversing the argument, the observed Stokes profiles 
of a single pixel provide a little amount of information to all modes simultaneously. This large redundancy is a very interesting advantage of 
our approach and introduces a strong regularization of the solution. This global character also introduces strong regularization 
due to the presence of spatial correlation in the observables and the physical parameters.
\end{itemize}

The main problem of the method that we present here is that a suitable linear transform has to
be chosen a-priori. The wavelet transform is sufficiently versatile so that it can highly compress the 2D maps of many of the physical parameters
needed for the synthesis of spectral lines. For this reason, our suggestion is to use the same transform $\mathbf{W}$ for all
the parameters. However, the method explained in the following allows the user to choose a different linear operator
$\mathbf{W}$ for every parameter $\mathbf{p}_i$. This way, it is possible to design a set of linear transforms that leads to a very sparse
representation of all the needed physical parameters. This additional step that we need to give as compared with standard
pixel-by-pixel inversions is the one that introduces the spatial regularization, though.

\subsection{Merit function}
All standard inversion codes work by optimizing a merit function that measures the 
$\ell_2$-norm of the residuals with respect to the vector $\mathbf{p}$ of physical parameters. 
If the observed Stokes parameters are corrupted with Gaussian noise with diagonal covariance
matrix, the merit function for the whole map is:
\begin{equation}
\chi^2_\mathbf{p} = \frac{1}{4N_\lambda N_\mathrm{pix}} \sum_{k=1}^{N_\mathrm{pix}} \sum_{i=1}^4 \sum_{j=1}^{N_\lambda} w_i \frac{\left[S_i(\lambda_j,\mathbf{\hat{p}}_k)-O_i(\lambda_j,k)\right]^2}{\sigma_{ijk}^2}
\end{equation}
where $\mathbf{S}(\lambda_j,\mathbf{\hat{p}}_k)=(I(\lambda_j,\mathbf{\hat{p}}_k),Q(\lambda_j,\mathbf{\hat{p}}_k),U(\lambda_j,\mathbf{\hat{p}}_k),V(\lambda_j,\mathbf{\hat{p}}_k))$ refers 
to the synthetic Stokes vector at wavelength 
position $j$ and position $k$. Likewise, 
$\mathbf{O}(\lambda_j,k)$ is the observed Stokes vector at this very same wavelength and for pixel $k$. The symbol $\sigma_{ijk}$ stands for the standard deviation of the noise
at wavelength position $i$, for the Stokes parameter $j$ and pixel $k$, while $w_i$ is the weight associated to
each Stokes parameter (that we assume, for simplicity, the same for all wavelengths and pixels). This weight is introduced for technical reasons to help improve the 
convergence during the optimization\footnote{Despite its technical interpretation, it can also be interpreted
as a modification of the noise variance associated to each Stokes parameter. It is customary to use the same noise
variance for all the Stokes parameters and then modify them using $w_i$.}. Classically, it is customary to solve the
problem 
\begin{equation}
\argmin_\mathbf{p} \, \chi^2_\mathbf{p},
\label{eq:problem_l2}
\end{equation}
by direct application of the Levenberg-Marquardt algorithm, which is specially suited to the optimization of such $\ell_2$-norms. This involves an
optimization for all parameters of all pixels simultaneously. Note that one can indistinguishably use $\mathbf{\hat{p}}$ or $\mathbf{p}$ because the
optimization has to be done for all parameters of all pixels.

In a complete parallel formulation, one can work on the transformed domain and write the merit function as:
\begin{equation}
\chi^2_\mathbf{q} = \frac{1}{4N_\lambda N_\mathrm{pix}} \sum_{k=1}^{N_\mathrm{pix}} \sum_{i=1}^4 \sum_{j=1}^{N_\lambda} w_i \frac{\left[S_i(\lambda_j,\left[\mathbf{W}^{-1}[\mathbf{q}]\right]_k)-O_i(\lambda_j,k)\right]^2}{\sigma_{ijk}^2}.
\end{equation}
The previous expression takes into account that the synthetic Stokes profiles at pixel $k$ depend now on the full vector $\mathbf{q}$ through the linear transformation.
Therefore, the inversion problem is solved by computing 
\begin{equation}
\argmin_\mathbf{q} \, \chi^2_\mathbf{q}.
\label{eq:problem_l2_q}
\end{equation}
The solutions to Eqs. (\ref{eq:problem_l2}) and (\ref{eq:problem_l2_q}) are strictly equivalent, with the difference that the former is solved
in the physical parameters, while the latter in the transformed domain. However, at the light of the 
results of Figs. \ref{fig:testSparsityQuiet}, \ref{fig:testSparsitySpot} and \ref{fig:testSparsityDiff}, Eq. (\ref{eq:problem_l2_q}) allows us to impose
the sparsity constrain, so that the regularized solution is obtained by computing:
\begin{equation}
\argmin_\mathbf{q} \chi^2_\mathbf{q},  \,\,
\mathrm{subject\, to\,} \Vert \mathbf{q} \Vert_0 \leq s,
\label{eq:problem_l0}
\end{equation}
where $\Vert \mathbf{q} \Vert_0$ is the $\ell_0$-norm of $\mathbf{q}$, equivalent to counting the number of non-zero elements. The desired sparsity
level is set by the upper limit $s \ll N_\mathrm{par} N_x N_y$. In other words, we minimize the merit function $\chi^2$ with respect to the 
modes $\mathbf{q}$ but using only a very small number of non-zero elements in $\mathbf{q}$. Equivalently, the
sparsity constrain can be incorporated in the function through a Lagrange multiplier, so that Eq. (\ref{eq:problem_l0}) can be rewritten as:
\begin{equation}
\argmin_\mathbf{q} \chi^2_\mathbf{q} + \lambda \Vert \mathbf{q} \Vert_0.
\label{eq:problem_l0_synthesis}
\end{equation}
with $\lambda$ a parameter that controls the strength of the constraint.

\subsection{Proximal algorithms}
As a consequence of the recent huge increase on the interest of compressed sensing, matrix completion, big data and related techniques
based on the idea of sparsity, several algorithms have been developed to optimize the addition of convex functions including non-convex 
constraints ($\ell_0$ or $\ell_1$ norms, for instance). The objective is to solve the following problem:
\begin{equation}
\argmin_\mathbf{q} f(\mathbf{q}) + g(\mathbf{q}),
\label{eq:general_proximal_problem}
\end{equation}
where $f(\mathbf{q})$ is a convex and differentiable function, while $g(\mathbf{q})$ is another convex function, possible non-smooth.
This is exactly the type of problem that we have to solve in Eq. (\ref{eq:problem_l0_synthesis}), in our case with $f(\mathbf{q})=\chi^2_\mathbf{q}$
and $g(\mathbf{q})=\lambda \Vert \mathbf{q} \Vert_0$.

One of the most successful algorithms developed recently belong to the class of methods termed
\emph{proximal algorithms} \citep[e.g.,][]{parikh_boyd14}. They can be viewed as the equivalent of the
Newton method for non-smooth, constrained, and large-scale optimization problems. In this case, the solution 
to the general problem (\ref{eq:general_proximal_problem}) is given by the following iteration \cite[e.g.,][]{starck10}:
\begin{equation}
\mathbf{q}_{i+1} = \mathrm{prox}_g \left[ \mathbf{q}_i - \mathbf{h} \nabla_\mathbf{q} f(\mathbf{q}_i) \right],
\label{eq:solutionIteration}
\end{equation}
where one carries out a step along the gradient of $f(\mathbf{q})$ and then applies the so-called \emph{proximity operator}, defined in the following. 
Note that, since $f(\mathbf{q})=\chi^2_\mathbf{q}$ is differentiable, the term inside parenthesis is the standard gradient descent method that are typically used 
in many inversion codes for the Stokes parameters.

It is clear from Eq. (\ref{eq:solutionIteration}) that the key ingredient of these
proximal algorithms is the application of the \emph{proximity operator} of the regularization function $g(\mathbf{x})$, defined as \citep{parikh_boyd14}:
\begin{equation}
\mathrm{prox}_g(\mathbf{x}) = \argmin_\mathbf{v} \left[ \frac{1}{2} \Vert \mathbf{v}-\mathbf{x} \Vert_2^2 + \lambda g(\mathbf{v}) \right].
\end{equation}
In general, this proximal operator has to be computed numerically, but exact solutions exist for many interesting regularization terms \citep{parikh_boyd14}. In fact,
in our case that we use the $\ell_0$-norm as regularization, the proximity operator is simply given by:
\begin{equation}
\mathrm{prox}_{\ell_0}(\mathbf{x}) = H_s(\mathbf{x}),
\label{eq:proximal_l0}
\end{equation}
where $H_s(\mathbf{x})$ is the nonlinear hard thresholding operator. This operator keeps the $s$ elements of $\mathbf{x}$
with largest absolute value untouched and sets to zero the remaining elements. Although all the results presented in this paper
use the $\ell_0$-norm as regularization, sometimes it is interesting to use the $\ell_1$-norm\footnote{The $\ell_1$-norm
is given by: $\Vert \mathbf{x} \Vert_1 = \sum_i |x_i|$.}. In this case, the proximal operator is the smooth thresholding
operator, which is given by:
\begin{equation}
\mathrm{prox}_{\ell_1}(\mathbf{x}) = \mathrm{sign}(\mathbf{x}) (|\mathbf{x}|-\lambda)_+,
\end{equation}
where $(\cdot)_+$ denotes the positive part.


\begin{algorithm}[!t]
\KwData{Stokes profiles and model atmosphere}
\KwResult{$\ell_0$-regularized solution}
Initialization: $t_0=1$ and $\mathbf{y}_1=\mathbf{q}_0$, with $\mathbf{q}_0$ a first estimation of the modes\;
 \While{not converged}{
 $\mathbf{q}_{k} = H_s \left[ \mathbf{y}_k - \mathbf{h} \nabla_\mathbf{q} \chi^2(\mathbf{y}_k) \right]$\;
 $t_{k+1}= \frac{1+\sqrt{1+4t_k^2}}{2}$\;
 $\mathbf{y}_{k+1} = \mathbf{q}_{k} + \left( \frac{t_k-1}{t_{k+1}} \right) (\mathbf{q}_{k} - \mathbf{q}_{k-1})$\;
 \If{$\nabla_\mathbf{q} \chi^2(\mathbf{y}_k) (\mathbf{q}_{k} - \mathbf{q}_{k-1}) > 0$}{
   $\mathbf{q}_0=\mathbf{y}_0=\mathbf{x}_k$, $t_k=1$ and $k=0$\;   
 } 
 }
 \Return $\mathbf{q}_k$
 \caption{FISTA algorithm with restarting strategy.} 
 \label{alg:fista}
\end{algorithm}

Putting together Eqs. (\ref{eq:solutionIteration}) and (\ref{eq:proximal_l0}), the simplest proximal algorithm that uses the gradient descent method 
to solve Eq. (\ref{eq:problem_l0_synthesis}) is given by \cite[e.g.,][]{parikh_boyd14}:
\begin{equation}
\mathbf{q}_{k+1} = H_s \left[ \mathbf{q}_k - \mathbf{h}
\nabla_\mathbf{q} \chi^2(\mathbf{q}_k) \right].
\label{eq:fista}
\end{equation}
This iteration is just a plain gradient descent algorithm (which only makes
use of first order derivatives) which is augmented by using a hard 
thresholding projection operator in each iteration. In other words, after moving 
the solution on the direction of the negative gradient (controlled by the
step-size $\mathbf{h}$), one sets to zero all elements that do not fulfill the 
sparsity constraint. This simple iterative scheme displays a sublinear convergence
rate $O(1/k)$. Recently, a trivial improvement of the gradient descent method
known as Fast Iterative Shrinkage-Thresholding Algorithm (FISTA)
has been developed by \cite{beck_teboulle09} showing a quadratic convergence rate $O(1/k^2)$. Algorithm
\ref{alg:fista}, a combination of FISTA and the restarting scheme developed by \cite{odonogue12}, is our
method of choice in this paper. We leave for the future the analysis of algorithms that smartly 
introduce some second-order information without ever constructing the Hessian matrix \citep[e.g.,][]{becker12}.

Eq. (\ref{eq:fista}) shows that the optimization method relies on the computation of the gradient of the merit function with respect
to the modes $\mathbf{q}$, that we have made explicity by using the subindex $\mathbf{q}$ on $\nabla_\mathbf{q} \chi^2$. Given
the linear character of the transformation of Eq. (\ref{eq:linearTransformation}), this gradient can be 
trivially related to the gradient with respect to the physical parameters (the response functions), that can be 
analitically written as:
\begin{equation}
\nabla_{\mathbf{q}_i} \chi^2 = \mathbf{W}^{-1} \nabla_{\mathbf{p}_i} \chi^2, \qquad i=1,\ldots,N_\mathrm{par}.
\label{eq:transform_gradient}
\end{equation}

The method described in this section is independent of the specific details of the transform $\mathbf{W}$. The only condition is that it has to
be linear and invertible. Apart from that, it is also desirable from the computational point of view that the linear transform has a fast algorithm available.
This is the case both for the orthogonal wavelet and DCT transforms.

\subsection{Possible improvements}
Even though gradient descent methods are known to be fast when approaching the minimum, 
they become slower in the refinement phase of the solution (that is precisely the reason why the 
Levenberg-Marquardt method is a combination of a gradient descent method and a Newton method, controlled by the 
Hessian). However, the non-convex optimization of large-scale problems has to rely on first-order derivatives
because the calculation and storage of the Hessian is impractical. In our experience and that of many others in the
literature, the FISTA algorithm is a competitive optimization technique. To achieve a good convergence speed, the 
vector of step-sizes $\mathbf{h}$ has to be smartly chosen. Extensive experiments have convinced us that these steps can be kept
fixed and used in different data sets without any special impact on the convergence speed.
However, we are currently investigating the possibility of improving the convergence speed along two lines.
First, using approximations to the diagonal
Hessian matrix that can be efficiently computed using quasi-Newton methods. Quasi-Newton methods
update the Hessian matrix by analyzing successive gradient vectors using a generalization of the secant method. We anticipate
that the methods developed by \cite{becker12} or \cite{marjugi13} can be of help. Second, using conjugate
gradient methods.

We have noticed that convergence is greatly improved if the inversion
is started forcing a very sparse solution, which in practice is similar to
reducing the number of parameters in the first steps \citep[]{sir92}.
Once the model cannot be further improved, the inversion can be re-started assuming the
final target sparsity, and therefore, giving more freedom to the inversion in the final steps. In fact, we have noticed 
that in cases where the sparsity is assumed to be low, the inversion can get stuck when $\chi^2$ is still very high. We 
blame the FISTA algorithm with fixed $\mathbf{h}$ for this behaviour. We expect to ameliorate this behaviour with the future 
improvements that are described above.


\begin{figure*}
\centering
\includegraphics[width=0.82\textwidth]{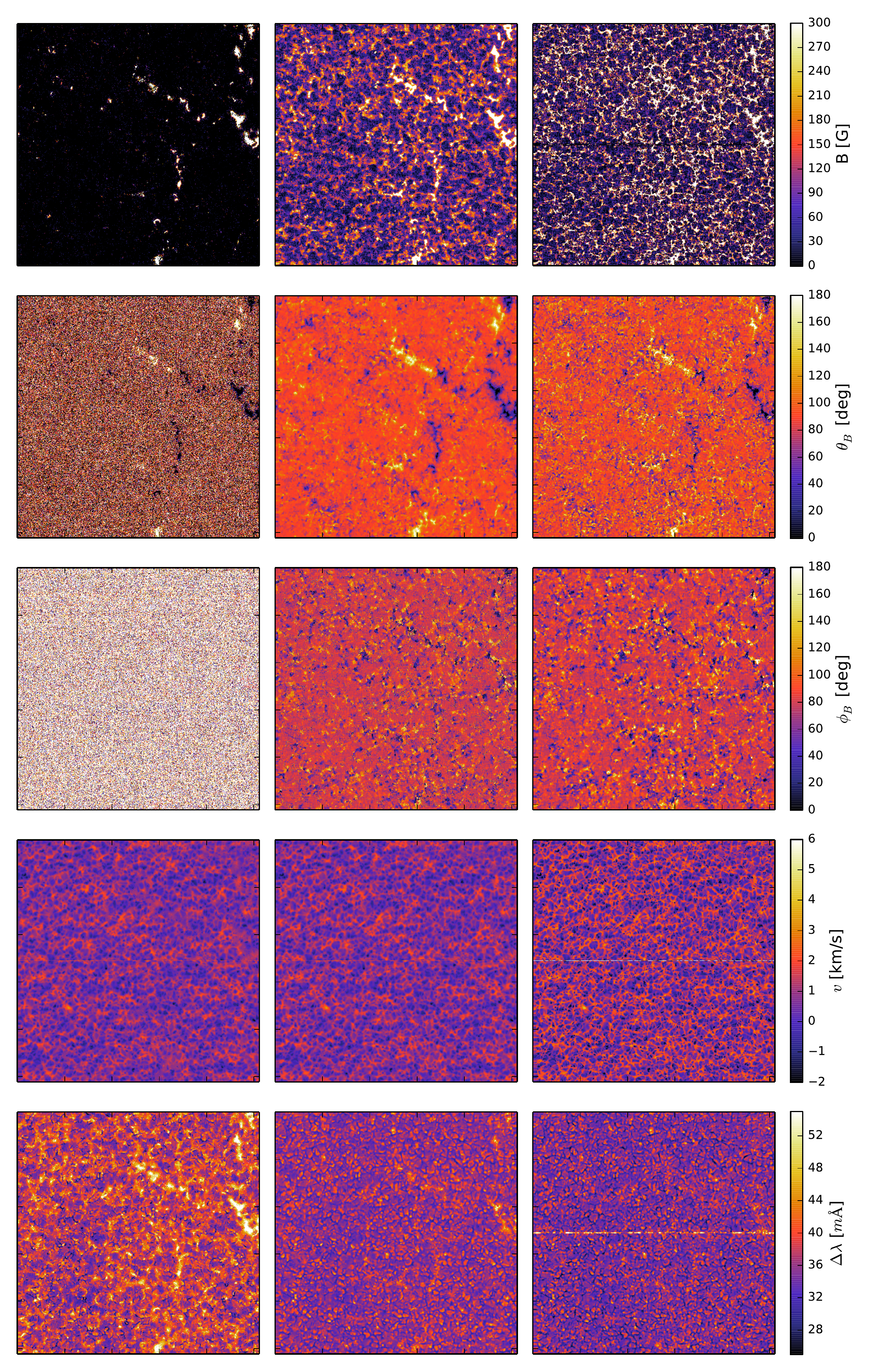}
\caption{Inverted maps for some of the most relevant physical parameters of the model for a quiet Sun observation. The map
is 512$\times$512 pixels. The left column shows the pixel-by-pixel results, while the central panels display
the results of the sparsity regularized inversion, using a Daubechies-8 orthogonal wavelet and a thresholding
value of 50\%. The third column corresponds to a sparsity regularized inversion compensating for the Hinode telescope PSF.}
\label{fig:resultQuiet}
\end{figure*}

\subsection{Summary}
The inversion scheme that we propose in this paper works as follows. A linear transform is chosen so that
the transformed 2D maps of model parameters are as sparse as possible. We have tested that DCT and wavelet
transforms seem to be appropriate. At the beginning of the iterative process, all the model parameters 
are initialized using any appropriate initialization like any standard inversion code. The initial solution
is transformed to obtain the first estimation of the modes.
At each iteration, the value of the merit function and of its gradient with 
respect to the model parameters taking into account all observed pixels are computed. To this end, the inverse
transform of the current solution has to be computed. The gradients, reordered 
as 2D maps, are transformed using Eq. (\ref{eq:transform_gradient}). Once the gradient descent step is applied, the sparsity 
constrain is applied by hard thresholding the resulting new modes. In this somehow trivial step is where the sparsity condition
is applied. Finally, algorithm \ref{alg:fista} is repeated until convergence.

\section{The computer code}
We have developed a computer code using the ideas presented in the previous section. The code is written
in C++ using the Message Passing Interface (MPI) standard for parallelization using a master-slave strategy. One
node works as a master, sending calculations to the slaves and receiving the results. Although the architecture of the
code is very general, the current version of the code only works with the ME model atmosphere. It can be used to
carry out pixel-by-pixel inversions (using the LM algorithm) and also sparsity regularized
two-dimensional inversions (using the FISTA algorithm). We will provide extensive details on the
implementation in a near future, once the code is expanded to use
more realistic cases than the ME included in this paper (de la
Cruz Rodr\'iguez et al. in prep.). Additionally, we will make the code
freely available.

\begin{figure}
\centering
\includegraphics[width=\columnwidth]{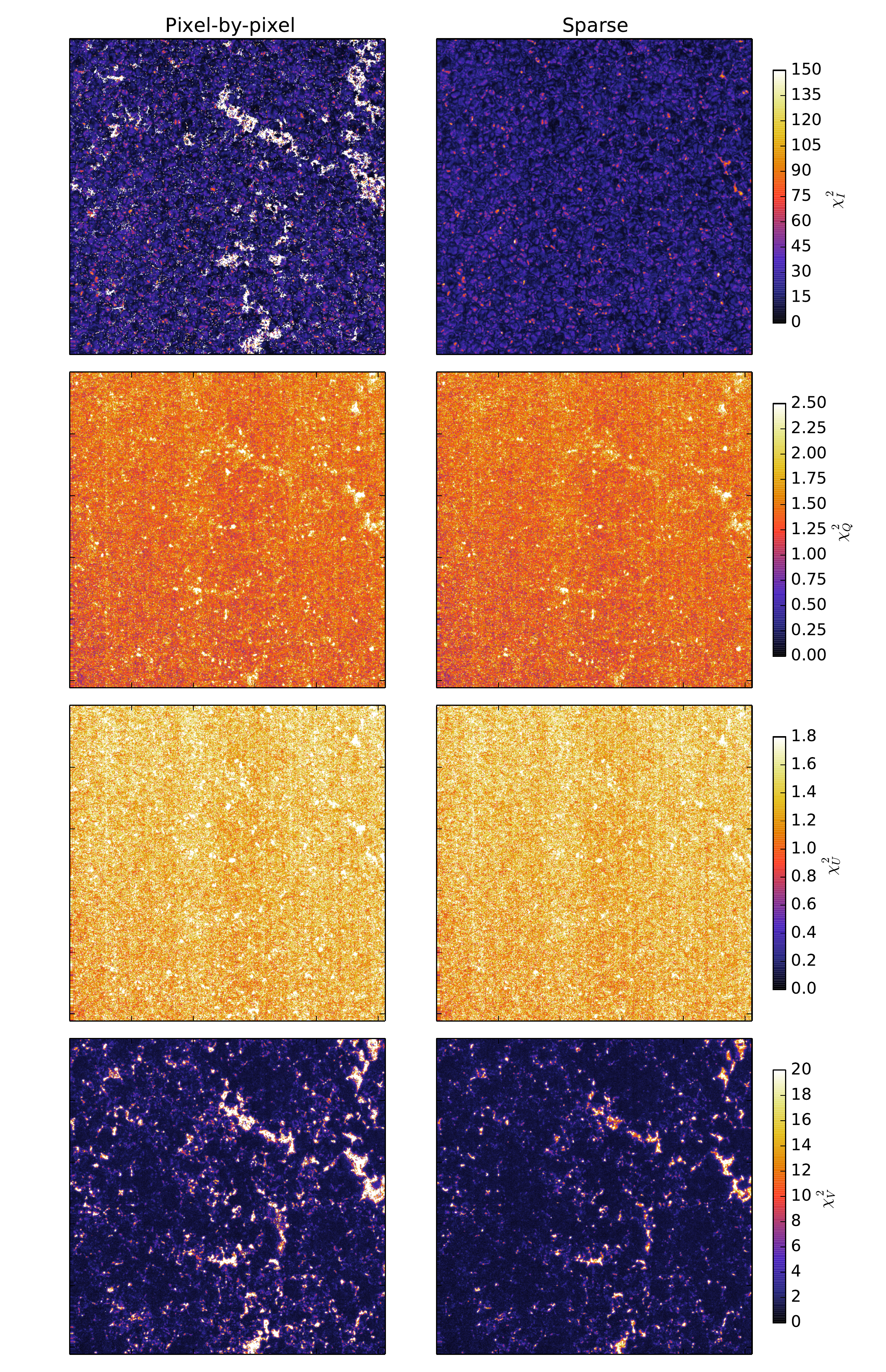}
\caption{Value of the $\chi^2$ for each Stokes parameter for the pixel-by-pixel inversion (left panels) and the sparsity regularized inversion (right panels).}
\label{fig:quietChi2}
\end{figure}

The code starts by using an initial estimation of the transformed coefficients for the maps of parameters. With this
initialization, the computationally heavy part of the calculation of the $\chi^2$ and the gradient term $\nabla_\mathbf{p} \chi^2$ is fully parallelized, with each
slave working on a piece of the map. When all the information is gathered by the master node, it computes the gradient term
in the transformed domain by direct application of Eq. (\ref{eq:transform_gradient}). This are the needed quantities to
evaluate one iteration of the FISTA algorithm and produce a correction of the transformed coefficients of the 
maps of physical quantities. This very same procedure is iterated until convergence.

The code also allows the user to specify a PSF that is used for degrading the synthetic profiles and carry out 
the spatial deconvolution and the regularized inversion simultaneously, similar to the approach followed by
\cite{vannoort12}. The inclusion of the deconvolution is straightforward in our first-order approach. The convolution operator is linear
so that, apart from convolving the Stokes profiles with the PSF, we only need to convolve the response functions with the PSF, so that Eq. (\ref{eq:transform_gradient}) becomes
\begin{equation}
\nabla_{\mathbf{q}_i} \chi^2 = \mathbf{P} \ast \left[\mathbf{W}^{-1} \nabla_{\mathbf{p}_i} \chi^2 \right], \qquad i=1,\ldots,N_\mathrm{par}
\label{eq:transform_gradient_convolution}
\end{equation}
where $\ast$ is the convolution operator.

\begin{figure*}
\centering
\includegraphics[width=0.82\textwidth]{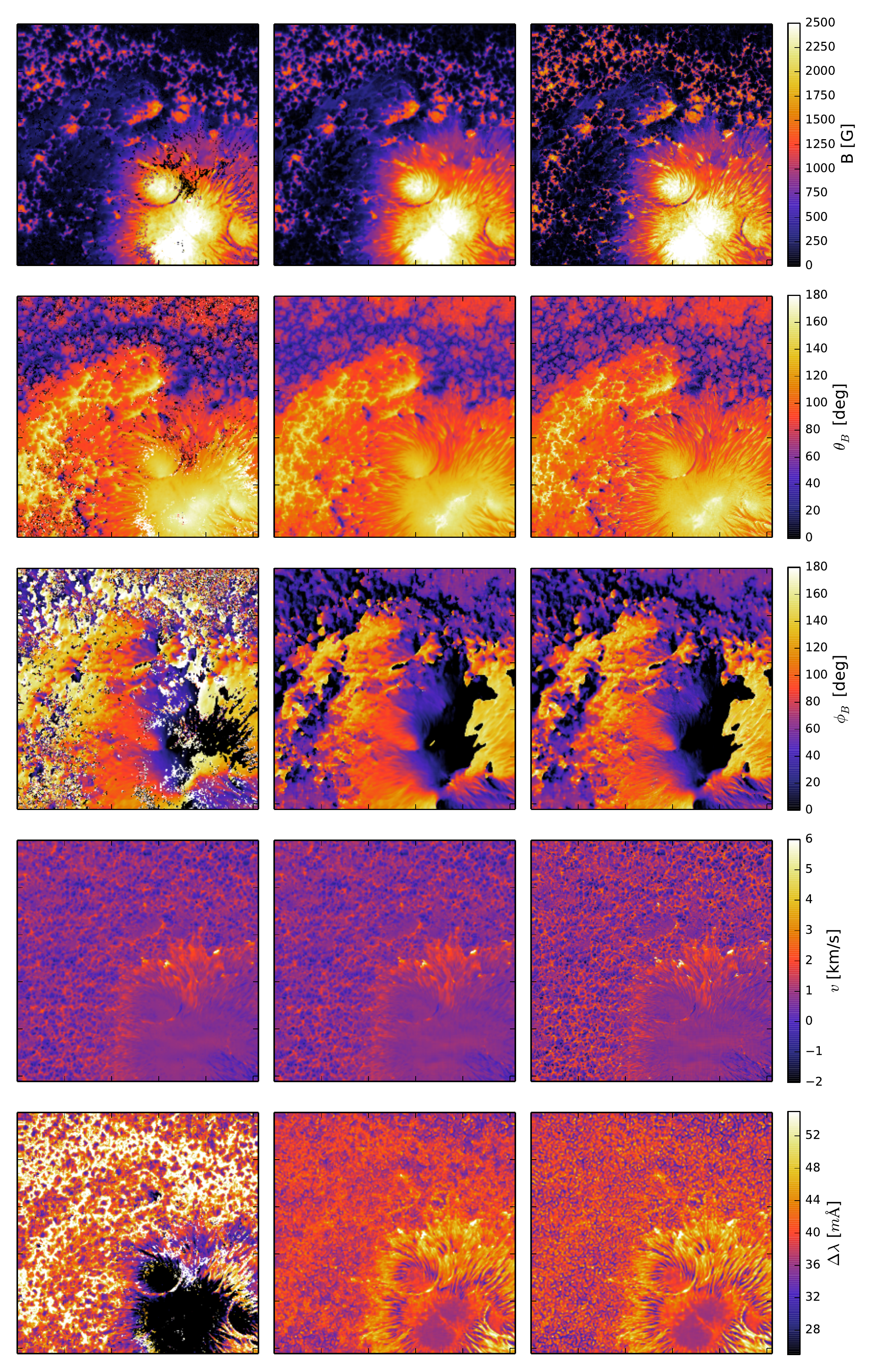}
\caption{Same as Fig. \ref{fig:resultQuiet} but for the sunspot data.}
\label{fig:resultSpot}
\end{figure*}

\section{Illustrative examples}
As a demonstration of our approach, we present in Fig. \ref{fig:resultQuiet} the inverted maps for some physical
parameters of the ME model in the quiet Sun map. Since the inversion is carried out with a very simple model
with only one component that is not able to capture the behavior of unresolved fields, we cannot infer much from
the magnetic properties on the map. However, a few things need to be commented:
\begin{enumerate}
\item The pixel-to-pixel inversion has been carried out in a multistep process to improve the quality
of the fit. First, we obtained a result from 50 randomizations of the initial parameters, keeping only the best solution for each pixel. 
Then we smooth the map with a Gaussian kernel. Finally, we re-run the inversion with another 50 randomizations, starting from the 
smoothed maps. We keep the solution that produces the best fit.
\item The general appearance of
the maps is much cleaner for the sparsity regularized maps, a consequence of the thresholding. This avoids arbitrary
pixel-to-pixel variations because some spatial coherence is forced. 
\item The map of Doppler velocities are
indistinguishable for the two inversions, meaning that this quantity is very easy to find in a pixel-by-pixel
inversion. 
\item The spatially regularized inversion returns lots of fields inclined by $\sim$90$^\circ$. This is
a direct consequence of the fact that many pixels have very low signal-to-noise ratio (SNR). The bias produced
by the noise on the transversal component of the magnetic field \citep{marian_dipolo12} makes that the fields systematically
appear very inclined. 
\item We also find that the Doppler width of the line is larger in the granules than in the
intergranules on the sparsity regularized solution. The opposite is found in the pixel-by-pixel inversion. This behavior
is compensated in the sparsity regularized solution by an small increase of the magnetic field.
\item The maps in the third column of Fig.~\ref{fig:resultQuiet} show the effect of compensating for the telescope PSF. Features 
become more conspicuous and compact, as demonstrated by \citet{vannoort12} and \citet{ruizcobo_asensioramos13}.
\end{enumerate}

The quality of the fits is assessed by computing the associated $\chi^2$ for each Stokes parameter and shown in 
Fig. \ref{fig:quietChi2}. Curiously, the sparsity regularized inversion is comparable for Stokes $Q$, $U$ and $V$ and
much better for Stokes $I$, because the sparse solution does not show all the structuring visible in the pixel-by-pixel one.
This demonstrates that our inversion code finds a much simpler model (in terms of the number of free parameters)
that explains the data better.

Figure \ref{fig:resultSpot} shows the results for the sunspot data. Again, the inferred Doppler velocities are very similar
in the two approaches, giving robustness to this quantity. The same behavior is also seen in the Doppler width of the line, with
granules presenting spectral lines with an enhanced broadening as compared with the intergranular lanes. The same applies to the
penumbra, where we find dark filaments to have a smaller Doppler width. Concerning the magnetic field vector, we find smoother
smoother and much less noisy results.

\begin{figure}
\centering
\includegraphics[width=\columnwidth]{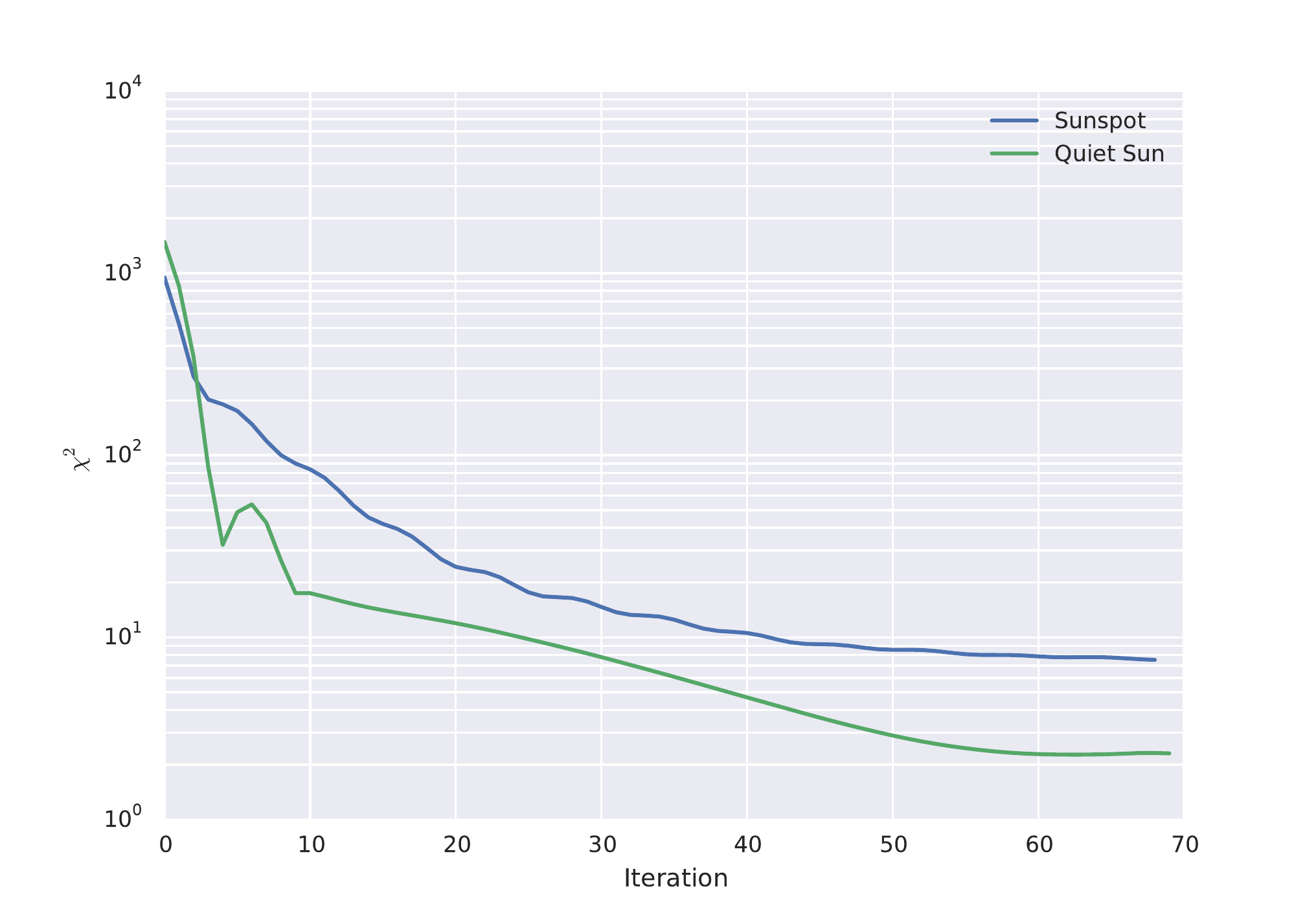}
\caption{Convergence of $\chi^2$ as a function of iteration number for the two data sets.}
\label{fig:convergence}
\end{figure}

Concerning the convergence speed, Fig. \ref{fig:convergence} displays the value of the reduced $\chi^2$ in the whole
map and taking into account all Stokes parameters for the two datasets. As usual in first-order, the initial convergence is
very fast, while the refinement of the solution is slower. We note that artifacts can appear in the sparse inversion 
when the solution is not fully converged. Unlike the pixel-by-pixel case, these artifacts appear as spatially coherent patches 
in the solution, that deviate from the surroundings. When the solution is converged, these patches disappear. Therefore, 
a different intuition is needed to assess whether the solution is physically meaningful or not.
  
\begin{figure*}
\centering
\includegraphics[width=\textwidth]{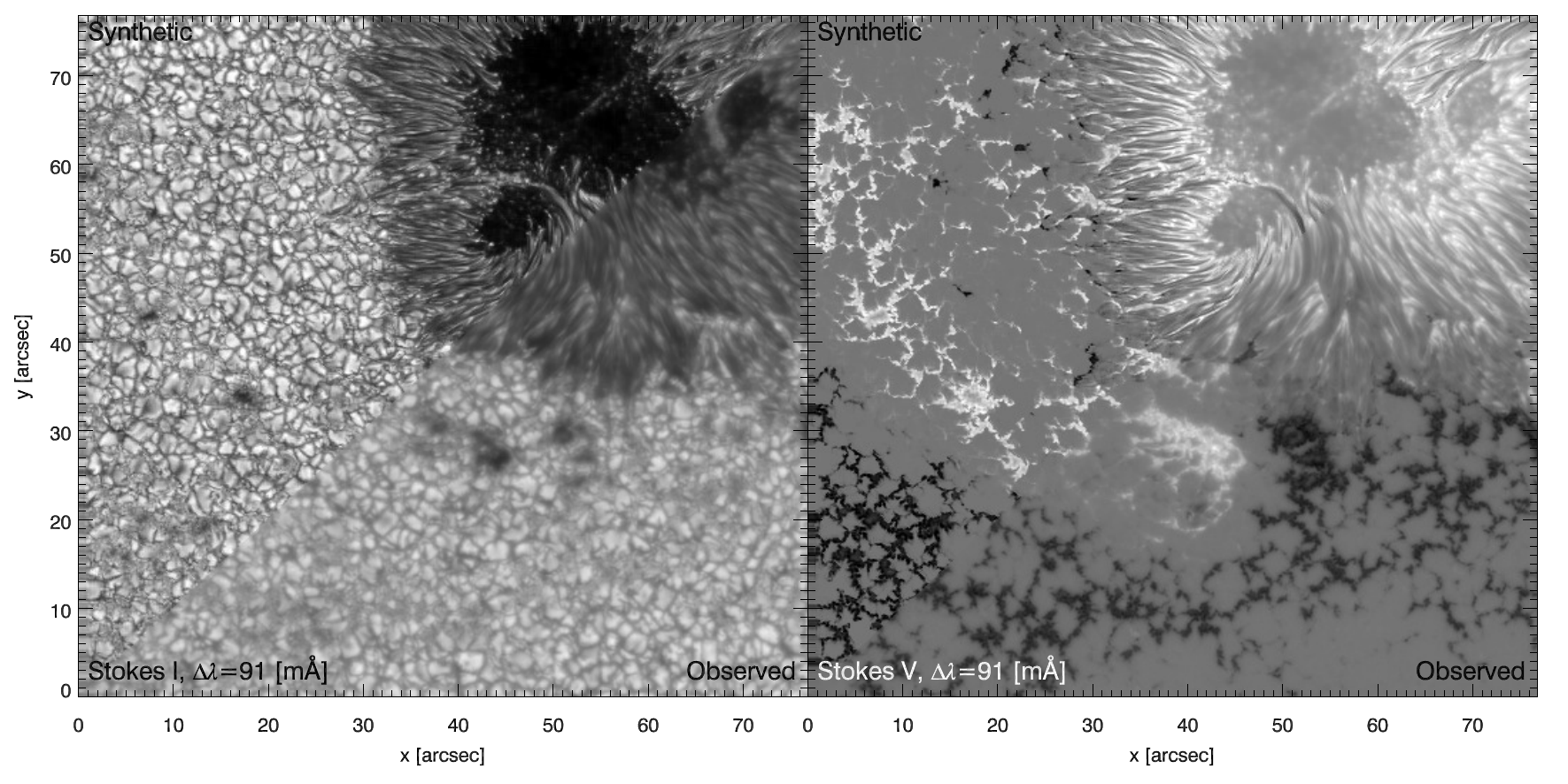}
\caption{Illustration of the effect of the telescope PSF in the
data. The FOV is divided in two halfs for a monochromatic image at
$\Delta\lambda=91$~m\AA\ from line center in Stokes~$I$ (left) and
$V$ (right). The lower half shows the observations, without applying
the PSF compensation. The upper half shows the synthetic image from
the inversion, including the PSF compensation. This inversion has
been performed assuming $50\%$ sparsity.}
\label{fig:spatpsf}
\end{figure*}

Finally, Fig.~\ref{fig:spatpsf} illustrates the smearing effect of the PSF in the observed data. We
have not noticed the oscillatory behavior that was reported by
\cite{vannoort12}. We speculate that the regularization present in our sparse
inversion naturally damps that behavior, but to be fair a ME model is
simpler than the depth-dependent case consider by \cite{vannoort12}
and this issue should be addressed under the same conditions.

\section{Conclusion and outlook for the future}
This paper presents the first full 2D regularized inversion code for Stokes profiles 
based on the idea of sparsity. Although the atmosphere model
that we use is very simple, it represents the first step toward a full
suite of inversion codes. The strong spatial regularization that we
introduce thanks to the sparsity constrain allows us to obtain very
reliable results. 

Our method is built upon some of the ideas presented by
\cite{vannoort12} and \cite{ruizcobo_asensioramos13}: the model is
spatially coupled and regularized. Furthermore, given that we do not
use a Hessian based inversion method, compensating the data for
telescope or stray-light PSF is trivial, as it does not add extra
complexity to the code, beyond the actual computation of
convolutions. However, if the telescope PSF is not known, the sparse
regularization will also couple the parameters from neighboring
pixels. The intrinsic nature of the wavelet transform allows to
reproduce small-scale and localized features, as well as the large
scales and therefore, reducing the number of free parameters does not
necessarily preclude the code from reproducing gradients and sharp
features if the correct sparsity level is chosen.

In this study we have used the first-order accelerated FISTA optimization algorithm to perform the 2D sparse
inversion. We emphasize that we are exploring alternative methods to further 
improve the convergence of the inversion.

A fast and parallel inversion code implementing everything
that is presented in this work and some extensions as described in the following will be freely available for the community soon. The code
will be extensible if the user provides the emergent Stokes profiles from a given parameterized atmosphere and
the response function of the Stokes parameters to all parameters. This response function can be computed analytically 
or, in general, numerically.

The concept of sparse regularization can be extended to many other problems that we plan to analyze in this series of papers, not necessarily maintaining
the order. In a first step, we plan to extend the sparse regularization to atmospheres
with gradients in the vertical direction. This will be done by fixing a set of heights in a common optical depth scale
(for instance, the optical depth scale at 500 nm) that will play the role of nodes, in parallel to which is currently
done in standard inversion codes like SIR, Nicole or SPINOR. The physical parameters (temperature, velocity, magnetic field, etc.)
at each node will be assumed to be sparse in a wavelet basis and the strategy used in this paper will be applied to
fit the emergent Stokes profiles.

In a second step, the concept of nodes will be surpassed. Each physical parameter will be determined by a 3D cube that will be
assumed to be sparse in a three-dimensional basis (DCT and orthogonal wavelets are trivially extended to the three
dimensional case). Therefore, the inversion will proceed by inferring a reduced number of 
wavelet modes for each physical parameter. The regularization imposed in this 3D inversion is very strong because
the observation of a single pixel provides (probably little) information for constraining the depth stratification of all the pixels
in the cube.

In a third step, standard pixel-by-pixel inversions can be carried out using a sparse regularization in height. The concept
of nodes, that have been present in all inversion codes with depth stratification looses its meaning again. The
depth stratification of the physical parameters of interest will be expanded in a suitable dictionary of depth-dependent
functions (we use the term dictionary because the functions do not need, in principle, to be orthogonal) and the optimization will
select the smallest number of functions needed to reproduce the Stokes profiles.

Finally, we note that our strategy could potentially be extended to do regularized superresolution. We leave this analysis for the future because it
needs to be studied in detail.

\begin{acknowledgements}
Financial support by the Spanish Ministry of Economy and Competitiveness 
through projects AYA2010--18029 (Solar Magnetism and Astrophysical Spectropolarimetry) and Consolider-Ingenio 2010 CSD2009-00038 
are gratefully acknowledged. AAR also acknowledges financial support
through the Ram\'on y Cajal fellowships. 

JdlCR acknowledges financial support from the
\textsc{Chromobs} project funded by the Knut and Alice Wallenberg
Foundation.

Part of the computations
included in this paper were
performed on resources provided by the Swedish National Infrastructure
for Computing (SNIC) at the National Supercomputer Centre (Link\"oping
University), with project id \emph{snic2014-1-273}. 

This research has made use of NASA's
Astrophysics Data System Bibliographic Services.
\end{acknowledgements}


\end{document}